\documentclass[a4paper,12pt]{article}
\usepackage{hyperref}
\usepackage{subfigure}
\usepackage{ae}
\usepackage{mathrsfs}
\usepackage{graphicx}
\usepackage{bbm}
\usepackage{latexsym}
\usepackage{dsfont}
\usepackage{amsmath}
\usepackage[title,titletoc]{appendix}
\usepackage[super]{nth}
\usepackage{cancel}
\usepackage{amsmath,amssymb}
\usepackage{mathtools}
\usepackage{cite}

\newcommand{\mathsym}[1]{{}}

\newcommand{\qed}{\nobreak \ifvmode \relax \else \ifdim\lastskip<1.5em \hskip-\lastskip \hskip1.5em plus0em minus0.5em \fi \nobreak \vrule height0.75em width0.5em depth0.25em\fi}

\def\app#1#2{  \mathrel{    \setbox0=\hbox{$#1\sim$}    \setbox2=\hbox{      \rlap{\hbox{$#1\propto$}}      \lower1.1\ht0\box0    }    \raise0.25\ht2\box2  }}

\begin{document}

\begin{titlepage} 
 \begin{center} \hfill CFTP\\

 \vskip 1cm {\large  \textbf{Do the Small Numbers in the Quark Mixing\\ arise from New Physics?}}

  G.C. Branco\footnote{gbranco@tecnico.ulisboa.pt},
José Filipe Bastos \footnote{jose.bastos@tecnico.ulisboa.pt} and
J.I. Silva-Marcos\footnote{juca@cftp.tecnico.ulisboa.pt}

 \vskip 0.07in Centro de
F{\'\i}sica Te\'orica de Part{\'\i}culas, CFTP, \\ Departamento de
F\'{\i}sica,\\ {\it Instituto Superior T\'ecnico, Universidade de Lisboa, }
\\ {\it Avenida Rovisco Pais nr. 1, 1049-001 Lisboa, Portugal} \end{center}
\begin{abstract}

We put forward the conjecture that the small numbers in the $V_\text{CKM}$ matrix, are generated by physics beyond the Standard Model. We identify as small numbers $V_{ub}$ and the strength of CP violation, measured by $|\text{Im}Q|$, where $Q$ stands for a rephasing invariant quartet of $V_\text{CKM}$. We illustrate how the conjecture can be realised in the context of an extension of the Standard Model where an up-type vector-like quark is introduced leading
to a realistic spectrum of quark masses and an effective $V_\text{CKM}$ in agreement with experiment. 

 \end{abstract}
\end{titlepage}

\section{Introduction}

In the Standard Model (SM) the pattern of fermion masses and mixing is
dictated by the flavour structure of Yukawa couplings. This flavour
structure is not constrained by gauge invariance and, as a result,
understanding the pattern of fermion masses and mixing in the SM continues
being at present an open, fundamental question. In a bottom-up approach to
the Yukawa puzzle, one usually tries to infer about the possible presence of
a family symmetry, from the observed pattern of quark masses and mixing.
This may be a specially difficult task if New Physics also contributes to the effective quark mass matrices at low energies.

In this paper, we put forward the conjecture that the small numbers in $V_%
\text{CKM}$ namely $|V_{ub}|$ and $I_\text{CP}=|\text{Im}Q|$ ($Q$ stands for a rephasing
invariant quartet of $V_\text{CKM}$) arise from New Physics.

The paper is organised as follows: in the next section we explain in detail
how to implement our conjecture concerning the origin of the small numbers
in $V_\text{CKM}$. In section 3, we analyse some of the phenomenological consequences of the model, in the
particular the mass of the new heavy-top quark and the size of
flavour-changing-neutral-currents (FCNC). Finally, we present a specific realistic model with
a vector-like up-type quark that realises our conjecture.

\section{The origin of the smallness of $|V_{ub}|$ and CP violation}

Quark mixing within the framework of the Standard Model (SM) is generated by
the Yukawa couplings which lead to quark mass matrices $M_dM^\dagger_d$, $%
M_uM^\dagger_u$, with different flavour structures. As a result, these two
matrices are diagonalised by two different unitary matrices, usually denoted 
$V^d_L$, $V^u_L$. Physically, only $V_\text{CKM} = V^{u\dagger}_LV^d_L$ is
measured. For a review see \cite{Branco:1999fs}. We have described the scenario of quark mixing in the SM. But the
SM leaves open many fundamental questions, so there is motivation to
consider New Physics (NP) beyond the SM. Then it is likely that NP also
contributes to the flavour structure of the effective quark mass matrices at
low energies, leading to an effective CKM which also reflects the presence
of New Physics.

As mentioned in the introduction, in this paper we put forward the
conjecture that the small numbers in the CKM matrix are generated by NP. We
adopt the standard parametrization of the CKM matrix \cite{Workman:2022ynf,Zyla:2020zbs} and identify as the
small numbers $|V_{ub}|\approx 4\times 10^{-3}$, and $%
I_\text{CP}=|\text{Im}Q|\equiv|\text{Im}\left(
V_{us}V_{cb}V_{ub}^{*}V_{cs}^{*}\right)|\simeq 3\times 10^{-5}$. We further
propose that there is a weak basis (WB) where the dominant contributions to
the orthogonal matrices $O_{12}$ and $O_{23}$ entering in the standard
parametrization arise from the down and up sectors, respectively. So
altogether we have

\begin{equation}
V_\text{CKM}^\text{eff}= \begin{pmatrix} 1 & 0 & 0 \\ 0 & c_{23} & s_{23} \\
0 & -s_{23} & c_{23} \end{pmatrix}_\text{up} \cdot \begin{pmatrix} c_{13} &
0 & s_{13} e^{-i\delta } \\ 0 & 1 & 0 \\ -s_{13} e^{i\delta } & 0 & c_{12}
\end{pmatrix}_\text{NP} \cdot \begin{pmatrix} c_{12} & s_{12} & 0 \\ -s_{12}
& c_{13} & 0 \\ 0 & 0 & 1 \end{pmatrix}_\text{down}  \label{vckm}
\end{equation}
where the subscripts indicate the origin of the contributions to the mixing
and NP stands for the contribution coming from New Physics. In the limit
where NP is not present, both $|V_{ub}|$ and $\text{Im}Q$ vanish.

\subsection{Framework}

In order to implement the structure in Eq. (\ref{vckm}), we assume that
there is a basis where the down and up quark matrices take the form: 
\begin{equation}
\begin{array}{lll}
M_{d}=\left( 
\begin{array}{lll}
m_{11}^{d} & m_{12}^{d} & 0 \\ 
m_{21}^{d} & m_{22}^{d} & 0 \\ 
0 & 0 & m_{33}^{d}
\end{array}
\right) & \quad \quad & M_{u}=\left( 
\begin{array}{lll}
m_{11}^{u} & 0 & 0 \\ 
0 & m_{22}^{u} & m_{23}^{u} \\ 
0 & m_{32}^{u} & m_{33}^{u}
\end{array}
\right)
\end{array}.
\label{mdu}
\end{equation}

Without the introduction of New Physics, one simply obtains a simplified and
reduced CKM mixing, where 
\begin{equation}
V_{\text{CKM}}=\begin{pmatrix} 1 & 0 & 0 \\ 0 & c_{23} & s_{23} \\ 0 &
-s_{23} & c_{23} \end{pmatrix}_{\text{up}}\cdot \begin{pmatrix} c_{12} &
s_{12} & 0 \\ -s_{12} & c_{12} & 0 \\ 0 & 0 & 1 \end{pmatrix} _{\text{down}}=%
\begin{pmatrix} c_{12} & s_{12} & 0 \\ -s_{12} c_{23} & c_{23} c_{12} &
-s_{23} \\ -s_{23} s_{12} & s_{23}c_{12} & c_{23} \end{pmatrix}.  \label{v}
\end{equation}

Note however that this limit already indicates that $\left| V_{31}\right|
=\left| V_{12}\right| \left| V_{23}\right| $ is much larger than $\left|
V_{13}\right| $, which vanishes at this stage, and that $V_{13}=0$ also leads to
vanishing CP violation.

The search for meaningful relations between the different CKM matrix elements (and/or mass ratios) has been part of the quest and the relentless effort of trying to unravel the flavour puzzle. For instance, recently Grossman and Ruderman studied in \cite{Grossman:2020qrp} the hypothesis of whether the CKM matrix has a substructure "that goes beyond the single small parameter of the Wolfenstein parameterization".

In the case studied here, it is important to notice, that our specific limit already points to a realistic
physical content. By this we mean that, if on the contrary, we
were to interchange the structure forms of $M_{u}$ and $M_{d}$ in Eq. (\ref
{mdu}), then we would obtain the result where $\left|
V_{13}\right| =\left| V_{12}\right| \left| V_{23}\right| $ and $V_{31}=0$,
which would clearly be in contradiction with experiment.
Indeed, when arguing that (some of) the elements of $V_\text{CKM}$ might be connected by some kind of relationship, what we consider most striking is that experimentally one already finds that $\left| V_{31}\right|
\approx \left| V_{12}\right| \left| V_{23}\right| $, and this is exactly what occurs in models where the substructures in Eq. (\ref{mdu}) are satisfied.

Now, when one introduces New Physics and extends the up quark sector with one isosinglet
VLQ, one obtains a $4\times 4$ extended up-quark mass matrix with new
elements. Let us then assume that this new mass matrix structure is near to $%
M_{u}$ in Eq. (\ref{mdu}) and given by

\begin{equation}
\mathcal{M}_{u}=\begin{pmatrix} 0 & 0 & 0 & m_{14} \\ 0 & m_{22} & m_{23} &
m_{24} e^{i\beta} \\ 0 & m_{32} e^{i\alpha } & m_{33} & 0 \\ m_{41} & 0 &
- m_{43} e^{i\delta } & M \end{pmatrix},  \label{muu}
\end{equation}
where the $(m_{ij},M)$ are real, $\alpha,\beta,\delta\in[0,2\pi]$ and for simplicity, we have taken $%
m_{11}^{u}=0$ and $m_{34}^{u}=0$. In addition one may also choose a WB where $%
m_{42}^{u}=0$. 
These structures for the up and down quark mass matrices may
also be obtained by imposing a discrete $Z_{4}$ symmetry on the Lagrangian
(see Appendix \ref{app_A}).

However, the crucial point here is that, in this new scenario, $\left| V_{13}\right| $
is now effectively different from zero, and generated by the mixing with the heavy extra
vector-like particle. Consequently, one may now have significant CP violation. To see
this, it is useful to re-write $\mathcal{M}_{u}$ in a different WB where
some right-handed fields have been transformed, such as to obtain $\left(\mathcal{M}'_u\right)_{4i}=0$, $i=1,2,3$. One finds, in leading order 
\begin{equation}
\begin{array}{c}
\mathcal{M}_{u}\xrightarrow[]{\text{WB}}\mathcal{M}_{u}^{\prime }=\mathcal{M}%
_{u}\cdot \mathcal{W}_{u} \\ 
\\ 
\mathcal{M}_{u}^{\prime }=\begin{pmatrix} -\frac{m_{14} m_{41}}{M} & 0 &
\frac{m_{14} m_{43}}{M} e^{i\delta } & m_{14} \\ -\frac{m_{24}
m_{41}}{M}e^{i\beta} & m_{22} & m_{23} & m_{24}e^{i \beta} \\ 0 & m_{32} e^{i\alpha } &
m_{33} & -\frac{m_{33 }m_{43}}{M}e^{-i\delta }\\ 0 & 0 & 0 & M \end{pmatrix}
\end{array}
\label{muuu}
\end{equation}
where we assume that $\left| m_{ij}\right| \leq \left| m_{33}\right| \ll
\left| M\right| $.

Several results can then be, immediately, derived from Eq. (\ref{muuu}). The
first one is that the effective up-quark mass matrix for the three lightest quarks is
given by 
\begin{equation}
M_{u}^{\text{eff}}=
\begin{pmatrix}
-\frac{m_{41} m_{14}}{M} & 0 & \frac{m_{14}m_{43}}{M}e^{i\delta } \\ 
-\frac{m_{41} m_{24}}{M}e^{i\beta } & m_{22} & m_{23} \\ 
0 & m_{32}e^{i\alpha } & m_{33}
\end{pmatrix}.   \label{ef3}
\end{equation}
Then, taking here the limit $\ m_{41}=0$, corresponding to $m_{u}=0$, and
assuming that $M\approx m_{T}$, the mass of the heavy vector-like quark, and 
$m_{33}\approx m_{t}$ while the other parameters are (much) smaller, one
already finds, in a rough approximation, for the diagonalization matrix $%
V^{u}$ of the up-quarks that 
\begin{equation}
V_{23}^{u}\approx \frac{m_{23}}{m_{t}},\quad \quad V_{13}^{u}\approx \frac{m_{14}m_{43}}{%
m_{T}m_{t}}e^{i\delta }.  \label{ro}
\end{equation}

From from Eq. (\ref{muuu}), one also obtains rough estimates for the extra
mixing angles involving the New Physics coming from the extra heavy up
quark, the heavy top. It easy to see that 
\begin{equation}
V_{14}^{u}\approx \frac{m_{14}}{m_{T}},\quad \quad V_{24}^{u}\approx \frac{m_{24}}{m_{T}}%
,\quad \quad V_{34}^{u}\approx- \frac{m_{t}m_{43}}{m_{T}^{2}}e^{-i\delta }.
\label{xtra}
\end{equation}

These estimates correspond to what one also finds when doing a rigorous 
calculation of the $4\times 4$ unitary matrix $\mathcal{V}^{u}$ which diagonalizes  $\mathcal{M}_{u}\mathcal{M}_{u}^\dagger$.

Finally, taking into account the $O_{12}$ mixing coming from the down quarks\footnote{Here for convenience, we embed the $O_{12}$ in a $4\times 4$ context}, and this  
$\mathcal{V}^{u}$, we obtain for the full quark mixing
$\mathcal{V}=\mathcal{V}^{u\dagger} O_{12}$,
where in leading order 
\begin{equation}
\mathcal{V}\approx \begin{pmatrix} c_{12} & s_{12} & \frac{m_{14}
m_{43}}{m_t m_T}e^{-i\delta} & \frac{m_{14}}{m_T}\\ \\ -s_{12} & c_{12} &
\frac{m_{23}}{m_t} & \frac{m_{24}}{m_T}e^{-i \beta} - \frac{m_{23}}{m_t} \frac{m_t m_{43}}{m^2_T}e^{i \delta} \\ \\
-c_{12}\frac{m_{14}m_{43}}{m_tm_T}e^{i\delta}+s_{12}\frac{m_{23}}{m_t} &
-c_{12}\frac{m_{23}}{m_t} & 1 &
-\frac{m_{t}m_{43}}{m_T^2}e^{i\delta}\\ \\ -c_{12}\frac{m_{14}}{m_T} &
-c_{12}\frac{m_{24}}{m_T}e^{i\beta}-s_{12}\frac{m_{14}}{m_T} & \frac{m_t
m_{43}}{m_T^2}e^{-i \delta} & 1 \end{pmatrix}.
\label{mixing}
\end{equation}
It is understood that the CKM mixing matrix $V_\text{CKM}$ is the $4\times 3
$ submatrix of (the left part of) $\mathcal{V}$. Note that the phase $\alpha$ plays no role at leading order. 

\subsection{The CKM Unitarity Problem and CP Violation }

Violation of $3\times 3$ unitarity is a proeminent feature of VLQ models, which in
turn makes them one of the most promising extensions of the SM when
addressing the CKM unitarity problem. 

Evidence for the CKM unitarity problem
stems from significant tensions between current determinations of $|V_{ud}|$
and $|V_{us}|$ and the assumption of $3\times 3$ unitarity of the SM \cite{Belfatto:2019swo,Seng:2018qru,Czarnecki:2019mwq,Seng:2020wjq,Hayen:2020cxh,Shiells:2020fqp,Czarnecki:2004cw,Coutinho:2019aiy,Aoki:2021kgd,Seng:2018yzq}.
Presently, the values for these mixings disfavour CKM unitarity at $2-3 \sigma $ and the best-fit value for the deviation to the unitarity of the
first row is $\Delta=0.0387\pm 0.0090$ \cite{Zyla:2020zbs}.  

With regard to our model, the deviation from unitarity is, primarily,
expressed in the normalisation of the first row of the CKM matrix, which
yields
\begin{equation}
|V_{ud}|^{2}+|V_{us}|^{2}+|V_{ub}|^{2}=1-\Delta^{2},
\end{equation}
where, in leading order, we have
\begin{equation}
\Delta=\left| \mathcal{V}_{14}\right| \approx \left| \frac{m_{14}}{m_{T}}%
\right| .  \label{v14}
\end{equation}

Now, when considering CP violation, induced by the mixing with the heavy
extra vector-like quark, and the CP-odd invariant quartet $|$Im$Q|\equiv |$Im$\left(
V_{us}V_{cb}V_{ub}^{*}V_{cs}^{*}\right) |$, we find, from Eq. (\ref{mixing}%
), that this invariant (in leading order) is given by
\begin{equation}
|\text{Im}Q|\approx \left| \mathcal{V}_{12}\right| \left| \mathcal{V}_{23}\right|
\left| \mathcal{V}_{13}\right| \left| \sin \delta \right|   \label{imq}
\end{equation}
where (again, in leading order)
\begin{equation}
\begin{array}{lll}
\left| \mathcal{V}_{12}\right| \simeq  s_{12} , & \left| \mathcal{V}%
_{23}\right| \approx \left|\frac{m_{23}}{m_{t}}\right|, & \left| \mathcal{V}_{13}\right| \approx \left| 
\frac{m_{43}}{m_{t}}\right| \left| \frac{m_{14}}{m_{T}}\right|. 
\end{array}
\label{v13}
\end{equation}
As one sees, from Eq. (\ref{v14}, \ref{imq}, \ref{v13}%
) in our model, the size of $\left| V_{ub}\right|=\left| \mathcal{V}_{13}\right| $, CP violation and the deviation from
unitarity in the first row of the CKM matrix are intrinsically connected. As an example, let us consider a unitarity deviation of $\Delta\approx0.04$, then with $\left| 
\frac{m_{43}}{m_{t}}\right|\approx0.1$, one obtains a value for $\left| \mathcal{V}_{13}\right|$ which is in agreement with experiment, $\left| \mathcal{V}_{13}\right|\approx0.004$, and with this we also recover the current measured value for $|\text{Im}Q|\approx 3 \times 10^{-5}$.

Thus, by proposing a common origin for both questions, our framework is able to explain (the smallness of) $V_{ub}$, and the CP violation observed in the quark sector and, simultaneously, provide a simple solution to the CKM unitarity problem.

\section{Phenomenology
\label{pheno}}

The introduction of a heavy-top implies changes to the SM electroweak Lagrangian, which will then lead to new contributions to various processes and consequently to the electroweak precision measurements (EWPM) associated with these. Some processes are stringently constrained by experiment and it is therefore crucial to analyse the respective phenomenological quantities.

With the addition of the extra quark, the mixing matrix is larger and the three standard down quarks now have an extra $W$-mediated mixing with the heavy-top. These contributions are relevant for processes such as $K^0-\overline{K}^0$ and $B_{d,s}^0-\overline{B}_{d,s}^0$ and correspond to a simple generalisation of the SM loop diagrams, but with the heavy-top also participating \cite{Aguilar-Saavedra:2002phh,Buras_2010,Cacciapaglia_2012}.  

Other contributions arise from flavour changing neutral current (FCNC) interactions with the $Z$ and
Higgs $h$ boson \cite{Botella_2017,Botella_2017_a,Nardi_1996,Barenboim:1997pf,Balaji:2021lpr}. In the physical basis, one finds for the neutral currents,

\begin{equation}
\begin{split}
\mathcal{L}_Z
=& \ {\frac{g}{c_W }} Z_{\mu } \left[ \frac{1}{2} \left( 
\overline{u}_{L}F^{u} \gamma ^{\mu }u_{L}- \overline{d}_{L}\gamma ^{\mu}d_{L}\right) -s^{2}_W \left( \frac{2}{3} \overline{u}\gamma ^{\mu }u-\frac{1}{3} \overline{d}\gamma ^{\mu }d\right) \right],\\ \\
-\mathcal{L}_{h}=& \ \overline{u}_L \frac{h}{v}F^u\mathcal{D}_u u_R+\overline{d}
_L \frac{h}{v}D_d d_R+\text{h.c.},  
\end{split}\label{VLQ_yukawa}
\end{equation}
where $\mathcal{D}_{u}$ stands for diagonal matrix containing all four up-type masses and $D_d$ the three
down quarks. The FCNC's are controlled by 
\begin{equation}
F^u=V_\text{CKM}V^\dagger_\text{CKM},
\end{equation}
which in general is non-diagonal.

In our framework, one obtains (in
leading order) for $F^u$ 
\begin{equation}
F^u\approx \begin{pmatrix} 1-\frac{m^2_{14}}{m^2_T} &
-\frac{m_{14}m_{24}}{m^2_T}e^{i\beta} & \frac{m_{14}m_{43}m_t}{m^3_T}e^{-i
\delta} & -\frac{m_{14}}{m_T}\\ \\ -\frac{m_{14}m_{24}}{m^2_T}e^{-i\beta} &
1-\frac{m^2_{24}}{m^2_T} & \frac{m_{24}m_{43}m_t}{m^3_T}e^{-i(\beta+\delta)}
& -\frac{m_{24}}{m_T}e^{-i \beta}\\ \\ \frac{m_{14}m_{43}m_t}{m^3_T}e^{i
\delta} & \frac{m_{24}m_{43}m_t}{m^3_T}e^{i(\beta+\delta)} & 1-
\frac{m^2_{43}m^2_t}{m^4_T} & \frac{m_{43}m_t}{m^2_T}e^{i\delta}\\ \\
-\frac{m_{14}}{m_T} & -\frac{m_{24}}{m_T}e^{i \beta} &
\frac{m_{43}m_t}{m^2_T}e^{-i \delta} & \frac{m^2_{14}}{m^2_T} \end{pmatrix}
\end{equation}
 which is indeed explicitly non-diagonal. This means that mixings between different up-type quark flavours are now allowed, leading, most notably, to NP contributions to $D^0-\overline{D}^0$ at tree level \cite{Branco_1995}. These dangerous effects are, nonetheless, suppressed by the size of the mass of the heavy-top $m_T$.
 
 Other processes, such as the golden modes $K_L\rightarrow \pi^0 \nu \overline{\nu}$ and $K^+\rightarrow \pi^+ \nu \overline{\nu}$ or the CP related parameter $\varepsilon^\prime/\varepsilon$ will also be affected by both types of contributions.

In the following, we briefly describe the viability of our model with respect to the most important EWPM's. In Appendix \ref{s14-dom}, we provide more detail.
An extensive analyses of the NP contributions to these phenomena was done in \cite{Branco:2021vhs,Botella:2021uxz,Belfatto:2021jhf}, where, in the context of an up-type VLQ, a solution to the CKM unitarity problem was proposed with $\Delta\approx 0.04$. 

For instance, the NP tree-level contribution to $D^0-\overline{D}^0$ is controlled by the size of $|F^u_{12}|$. Since, the SM contribution is negligible, one can use the upper-bound $x_D^\text{NP}< x_D^\text{exp}=0.39^{+0.11}_{-0.12}\%$ \cite{Golowich_2009,Amhis_2021} for the NP contribution to the mixing parameter $x_D$, which will lead to
\begin{equation}
    |F^u_{12}| < 1.22\times 10^{-3}.
\end{equation}
Taking e.g. a central value for $\Delta\simeq 0.0387$, one obtains an upper-bound for
\begin{equation}
    |\mathcal{V}_{24}| \lesssim 3.14\times 10^{-3}.
    \label{V24}
\end{equation}
Similarly, the ratio $m_{43}/m_T$ is constrained by the requirement\cite{Zyla:2020zbs}
\begin{equation}
    |\mathcal{V}_{13}|\simeq  |V_{ub}|^\text{exp}=\left(3.82\pm 0.20\right)\times 10^{-3},
\end{equation}
leading to $m_{43}\sim 15 \text{ GeV}$.
Moreover, currently, the least stringent lower-bound for the mass of the heavy-top is $m_T\simeq 0.685$ TeV, which originates from searches that assume dominant mixing of the heavy-top with the first generation \cite{Sirunyan_2018}. Using this result one has
\begin{equation}
    |\mathcal{V}_{34}|\simeq \frac{m_t m_{43}}{m^2_T}\lesssim 6.27 \times 10^{-3}.
    \label{V34}
\end{equation}
These upper-bounds for $|\mathcal{V}_{24}|$ and $|\mathcal{V}_{34}|$ are well within the allowed regions presented in \cite{Branco:2021vhs}, where the parameter regions of a general model with a heavy-top are scanned while imposing the experimental constraints coming from $K^0-\overline{K}^0$ and $B_{d,s}^0-\overline{B}_{d,s}^0$. Nonetheless, as it was pointed out in \cite{Botella:2021uxz}, the NP contributions to $\varepsilon_K$ are currently much more constrained, roughly $10\%$ of the standard model contribution which is very near to the experimental bound. Therefore, when introducing a VLQ, one should be very careful as not to exceed this bound. As it was shown, there exists a significant region of parameter space where the NP contribution to $\varepsilon_K$ are small (and $|\varepsilon^\text{NP}_K|\lesssim |\varepsilon^\text{exp}_K|/10$), and where
\begin{equation}
 |\mathcal{V}_{24}|, |\mathcal{V}_{34}|\ll|\mathcal{V}_{14}|\sim \lambda^2.
\end{equation}

In this region, the NP contributions to the golden mode kaon decays and $\varepsilon^\prime/\varepsilon$ are also within the experimental bounds (see Appendix \ref{s14-dom}).


\subsection{\textbf{Numerical Example}}
Next, we present a benchmark numerical example of our model and compute the NP contributions to the most relevant EWPM quantities.

We consider the following mass matrices (in GeV, at the $M_Z$ scale) for the down and up-sectors, respectively
\begin{equation}
\begin{array}{l}
M_{d}=\left( 
\begin{array}{ccc}
0.00292338 & -0.0134741 & 0 \\ 
0.000673705 & 0.0584675 & 0 \\ 
0 & 0 & 2.9
\end{array}
\right), 
\\
\\
\mathcal{M}_{u}=\left( 
\begin{array}{cccc}
0 & 0 & 0 & 53.7334 \\ 
0 & 0.59952 & -6.91815 & 1.250 e^{-0.285 i} \\ 
0 & -0.0239936 & 172.862 & 0 \\ 
0.046526 & 0 & 14.886 e^{-1.190 i} & 1250
\end{array}
\right).
 \label{m1}
\end{array}
\end{equation}

From these one obtains the following mass spectrum (also in GeV, at the $M_Z$ scale)  
\begin{equation}
\begin{array}{llll}
m_{d}=0.003, & m_{s}=0.060, & m_{b}=2.9, &
\\
\\
m_{u}=0.002, & m_{c}=0.60, & m_{t}=173, & m_T=1251.
\end{array}
\end{equation}
The CKM matrix is the $4\times 3$ left-sub-matrix of the following full $4\times 4$ mixing matrix
\begin{equation}
\left|\mathcal{V}\right|=\left( 
\begin{array}{cccc}
0.97354 & 0.224413 & 0.00370431 & 0.0429468 \\ 
0.224536 & 0.973644 & 0.0399975 & 0.000996211 \\ 
0.00833917 & 0.0393001 & 0.999192 & 0.00151171 \\ 
0.0416344 & 0.0105585 & 0.001674 & 0.999076
\end{array}
\right),  \label{ckm1}
\end{equation}
and the resulting CP violation rephasing invariant phases are
\begin{equation}
\begin{array}{l}
\gamma\equiv \text{arg}\left(-V_{ud}V_{cb}V^*_{ub}V^*_{cd}\right)
\simeq 68.0^\circ, \\ 
\\ 
\sin (2\beta )\equiv \sin\left[2 \ \text{arg}%
\left(-V_{cd}V_{tb}V^*_{cb}V^*_{td}\right)\right]\simeq 0.746, \\ 
\\ 
\chi \equiv \text{arg}\left(-V_{ts}V_{cb}V^*_{cs}V^*_{tb}\right)\simeq 0.020, \\ 
\\ 
\chi^\prime \equiv \text{arg}\left(-V_{cd}V_{us}V^*_{cs}V^*_{ud}\right)\simeq 5.71
\times 10^{-4}.
\end{array}
\label{cp1}
\end{equation}
with the CP-odd invariant quantity $I_\text{CP}=|\text{Im}Q|\equiv |\text{Im}\left(V_{ub}V_{cd}V^*_{ud}V^*_{cb}\right)|$,
\begin{equation}
   I_\text{CP}\simeq
3.00\times 10^{-5}.
\end{equation}

Finally, in table (\ref{t1}) we present the results for the most relevant EWPM quantities.

\begin{table*}[h]
\centering
\label{EWPMs} 
\begin{tabular}{|c|c||c|c|}
\hline
Observable & NP prediction & Observable & NP prediction \\ \hline
&  &  &  \\ 
$\Delta m_{B_d}$ & $1.50 \times 10^{-11} \text{ MeV}$ & $x_D$ & $0.048\%$ \\ 
&  &  &  \\ \hline
&  &  &  \\ 
$\Delta m_{B_s}$ & $2.79 \times 10^{-11} \text{ MeV}$ & $\varepsilon^\prime/%
\varepsilon$ & $-6.28 \times 10^{-5}$ \\ 
&  &  &  \\ \hline
&  &  &  \\ 
$\Delta m_{K}$ & $7.96 \times 10^{-13} \text{ MeV}$ & $\frac{\text{Br}%
\left(K^+\longrightarrow \pi^+ \overline{\nu}\nu\right)}{\text{Br}%
\left(K^+\longrightarrow \pi^+ \overline{\nu}\nu\right)_\text{SM}}$ & $0.429$
\\ 
&  &  &  \\ \hline
&  &  &  \\ 
$\varepsilon_K$ & $2.25 \times 10^{-5}$ & $\frac{\text{Br}%
\left(K^0\longrightarrow \pi^0 \overline{\nu}\nu\right)}{\text{Br}%
\left(K^0\longrightarrow \pi^0 \overline{\nu}\nu\right)_\text{SM}}$ & $0.636$
\\ 
&  &  &  \\ \hline
\end{tabular}
\caption{NP contributions to various processes for the numerical case in Eq.(\ref{m1}). }\label{t1}
\end{table*}

\section{Conclusions}

We have put forward the conjecture that the small numbers in $V_\text{CKM}$ originate from
Physics Beyond the SM. As small numbers, we identify $|V_{ub}|$ and the strength of 
CP violation, namely $I_\text{CP}= |\text{Im} Q|$, with $Q$ denotinq a rephasing invariant quartet of $V_\text{CKM}$.

We further propose that there is a weak basis where the effective $V_\text{CKM}$ matrix arises from a rotation in the $2-3$ up quark sector and a $1-2$ rotation in the down quark sector. 

Within this framework, in the SM both $V_{ub}$ and $I_\text{CP}$ vanish, and from quark mixing generated by SM Yukawa couplings, the exact relation $V_{31}= V_{12}V_{23}$ holds. In the framework of our conjecture one finds an explanation why $V_{31}$ is larger than $V_{13}$ and near to its experimental value. We propose a specific extension of the SM consisting of the addition of an up type vector like quark. It is also shown that the specific textures of Yukawa couplings result from a $Z_4$ symmetry imposed on the full Lagrangian.

\appendix
\section{Mass Matrix Structure with a $Z_{4}$ symmetry
\label{app_A}}

Bellow, we describe how to obtain the structures for the down quark mass matrix $M_d$ and the $4\times 4$ up-quark mass matrix $\mathcal{M}_{u}$ in a model with a discrete symmetry $Z_4$. To achieve this, we also introduce an extra scalar doublet $\phi_{2}$, and a singlet $S$. 

We give a transformation table for the left-hand fields $Q_{L_{i}}$,
the up and down-quark fields $u_{R_{i}},d_{R_{i}}$ and and their possible
couplings to the doublets $\phi$ , $\phi_{2}$, and the singlet $S$. These fields
transform as
\begin{equation}
\begin{array}{ccc}
\phi \longrightarrow \phi ,\quad & \phi _{2}\longrightarrow i\phi
_{2},\quad & S\longrightarrow -i S, \\ 
&  &  \\ 
\overline{Q}_{L1}\longrightarrow -\overline{Q}_{L1}, & \overline{Q}%
_{L2}\longrightarrow i\overline{Q}_{L2}, & \overline{Q}_{L3}\longrightarrow 
\overline{Q}_{L3}, \\ 
&  &  \\ 
U_{R,L}\longrightarrow i U_{R,L} & d_{R1}\longrightarrow -d_{R1}, & 
d_{R2}\longrightarrow -d_{R2},
\end{array}
\end{equation}
where the remaining RH quark fields ($u_{Rj}$ with $j=1,2,3$ and $d_{R3}$)
do not transform.

In the following transformation tables, we position next to each field, its
respective $Z_{4}$ property, and use zeros to denote a
forbidden entry (by the symmetry) in the quark mass matrices. 
\begin{table*}[h]
\centering
\label{symmetry_ups} 
\begin{tabular}{|c|cccc|}
\hline
Up-type & $u_{R1}$ & $u_{R2}$ & $u_{R3}$ & $U_{R}$ \\ 
quarks & $(1)$ & $(1)$ & $(1)$ & $(-i)$ \\ \hline
&  &  &  &  \\ 
$\overline{Q}_{L1} \ (-1)$ & $0$ & $0$ & $0$ & $y^u_{13} %
\Tilde{\phi}_2$ \\ 
&  &  &  &  \\ 
$\overline{Q}_{L2} \ (i) $ & $y^u_{21} \Tilde{\phi}_2$ & $y^u_{22} %
\Tilde{\phi}_2$ & $y^u_{23} \Tilde{\phi}_2$ & $y^u_{24} \Tilde{\phi}$ \\ 
&  &  &  &  \\ 
$\overline{Q}_{L3} \ (1)$ & $y^u_{31} \Tilde{\phi}$ & $y^u_{32} %
\Tilde{\phi}$ & $y^u_{33} \Tilde{\phi}$ & $0$ \\ 
&  &  &  &  \\ 
$\overline{U}_{L} \ (i)$ & $y^u_{41} S$ & $y^u_{42} S$ & $y^u_{43} S$ & $M$
\\ 
&  &  &  &  \\ \hline
\end{tabular}
\end{table*}
\begin{table*}[h]
\centering
\label{symmetry_ups} 
\begin{tabular}{|c|ccc|}
\hline
Down-type & $d_{R1}$ & $d_{R2}$ & $d_{R3}$ \\ 
quarks & $(-1)$ & $(-1)$ & $(1)$ \\ \hline
&  &  &  \\ 
$\overline{Q}_{L1} \ (-1) $ & $y^d_{11} \phi$ & $y^d_{12} \phi$ & $%
0 $ \\ 
&  &  &  \\ 
$\overline{Q}_{L2} \ (i) $ & $y^d_{21} \phi_2$ & $y^d_{22} \phi_2$ & $0$
\\ 
&  &  &  \\ 
$\overline{Q}_{L3} \ (1) $ & $0$ & $0$ & $y^d_{33} \phi$ \\ 
&  &  &  \\ \hline
\end{tabular}
\end{table*}

After a WB transformations of the up right-handed fields, we obtain the $4\times 4$ up-quark mass matrix $\mathcal{M}_{u}$ in Eq.(\ref{muu}), as well as the $3\times 3$ down-quark mass matrix in Eq.(\ref{mdu}).

\newpage

\section{Recovering the realistic $s_{14}$-dominance scenario
\label{s14-dom}}

Phenomenological constraints impose restrictions on the region of parameters and limits the size of the mixing of the heavy-top and the standard quarks. Our region of parameter space is such that
$ |\mathcal{V}_{24}|, |\mathcal{V}_{34}|\ll|\mathcal{V}_{14}|\sim \lambda^2$. 
More precisely, in terms of the mixing angles in the Botella-Chau parametrization \cite{Botella:1985gb}, we have in approximation
\begin{equation}
\begin{array}{ccc}
    s_{14}\sim \lambda^2, & s_{24}\lesssim \lambda^4, & s_{34}\lesssim \lambda^3.
\end{array}
\label{s14-dom-2}
\end{equation}
Comparing Eq. (\ref{mixing}) with the BC parametrization, in our framework one has that the angles and phases in the BC parametrization correspond, in approximation, to
\begin{equation}
    \begin{split}
    s_{14}\approx &\  \frac{m_{14}}{m_T}, \\   
    s_{24}\approx & \  \frac{m_{24}}{m_T},  \\  
    s_{34}\approx &\  \frac{m_t m_{43}}{m^2_T},
 \end{split}
\end{equation}

\begin{equation}
    \begin{array}{c}
    \delta'\equiv  \delta_{24}-\delta_{14}
\approx  \beta,\\
\\
    \delta_{14}\approx   \delta + \pi.
    \end{array}
\end{equation}
Only two phases are relevant at leading order, contrary to the three phases that typically participate in the general framework.

With this in mind, one finds that our model is compatible with the stringent experimental constraints coming from $K^+\rightarrow \pi^+ \nu \overline{\nu}$, $\varepsilon^\prime/\varepsilon$ and $\varepsilon_K$. 

With respect to $\varepsilon^\text{NP}_K$, it was argued \cite{Botella:2021uxz,Brod_2020}, that, since the SM-value and the experimental value of
 $|\varepsilon _{K}|$,
	\begin{equation}
	\begin{array}{ll}
		|\varepsilon ^{\text{SM}}_{K}|= (2.16\pm 0.18)\times 10^{-3},  \\
\\
|\varepsilon^{\text{exp}} _{K}|= (2.228\pm 0.011)\times 10^{-3},
	 \end{array}
\label{eksm}
	\end{equation}
are very close to each other, having  $	 |\varepsilon^{\text{exp}}_{K}|-|\varepsilon^{\text{SM}}_{K}|\simeq \left(0.68\pm 1.80\right)\times 10^{-4}$, the NP contribution to this EWPM parameter should be limited. At $1\sigma$ a new upper-bound for the NP contribution to $|\varepsilon _{K}|$ was established, such that $%
	|\varepsilon^{\text{NP}}_{K}|\lesssim 0.1|\varepsilon^{\text{exp}}_{K}|$, or more concretely
\begin{equation}
|\varepsilon^{\text{NP}}_{K}|^{1\sigma}\ \lesssim \Delta= 2.48\times 10^{-4},
\label{delta}
\end{equation}
	which severely restricts various models.
	
In the case studied here,
this crucial EWPM quantity now receives a new contribution, which was absent, in leading order, in the realistic $s_{14}$-dominance case presented in \cite{Botella:2021uxz}. We now have roughly (with the same phase convention) that, 
\begin{equation}
    |\varepsilon^\text{NP}_K|\propto s_{12}s^2_{14}\left|\eta^K_{tT} S(x_t,x_T)s_{13}s_{23}\sin\delta - \eta^K_{TT} S(x_T)s_{14}\left(s_{24}\sin\delta^\prime-s_{23}s_{34}\sin\delta\right)\right|,
    \label{epsilon_K_2}
\end{equation}
with the QCD corrections $\eta^K_{ij}$, the Inami-Lim functions $S(x_i,x_j)$, and $x_i\equiv m^2_i/M^2_W$ \cite{Inami:1980fz}. The expression for $\varepsilon^\text{NP}_K$ is very similar to the previously one obtained for the realistic $s_{14}$-dominance case.
The point is that, now we have an extra term proportional to $s_{34}$ and that within the region of parameters considered here, enough cancellation can also be easily achieved, allowing for small ($|\varepsilon^\text{NP}_K|\sim |\varepsilon^\text{exp}_K|/10$) contributions to $\varepsilon_K$.  
Thus, even here in this particular framework, we can still recover the main features of the realistic $s_{14}$-dominance case.

Furthermore, the NP contributions to $\text{Br}\left(K^+\rightarrow \pi^+ \nu \overline{\nu}\right)$ are controlled by $\lambda^K_T\equiv \mathcal{V}^*_{42}\mathcal{V}_{41}$,
In our case, we now have 
\begin{equation}
\lambda^K_T\approx s_{12}s^2_{14} +s_{14}s_{24}e^{i \delta^\prime},
\end{equation}
which has a similar form to the one obtained for realistic $s_{14}$-dominance case. Even, for values of $s_{24}$ closer to the upper-bound in Eq. (\ref{V24}), the first term will dominate while the second will still correspond to a smaller correction, as was the case in the realistic $s_{14}$-dominance scenario. Hence, one also finds that $\text{Br}\left(K^+\rightarrow \pi^+ \nu \overline{\nu}\right)$ is compatible with current experimental results. 

For the NP contributions to $\text{Br}\left(K_L\rightarrow \pi^0 \nu \overline{\nu}\right)$ and $\varepsilon^\prime/\varepsilon$, we have that these are controlled by $\text{Im}\left(\lambda^K_T\right)$,
\begin{equation}
\text{Im}\left(\lambda^K_T\right) \approx s_{14}\left(s_{24} \sin \delta^\prime - s_{23}s_{34}\sin\delta\right).
\end{equation}
where now we have also written the next order term.
This expression differs from the realistic $s_{14}$-dominance by a term in $s_{34}$, which is not necessarily negligible. Typically, this extra term will correspond to a correction to the dominant term with $s_{24}$, which was the only relevant term in the realistic $s_{14}$-dominance limit. However, for smaller values of $s_{24}$, we must have $\sin\delta^\prime\approx \sin \delta$ to ensure enough cancellation in (Eq. \ref{epsilon_K_2}), so that these terms can nearly cancel each other, and subsequently, also suppress $\left(\varepsilon^\prime/\varepsilon\right)_\text{NP}$ sufficiently enough. 

For the remaining golden mode, these changes are somewhat irrelevant as in our case case, $\text{Br}\left(K_L\rightarrow \pi^0 \nu \overline{\nu}\right)$ is of the same order as the SM prediction, which is still well below the current experimental bound \cite{Zyla:2020zbs}.

\section*{Acknowledgements}

This work was partially supported by Fundação para a Ciência e a Tecnologia (FCT, Portugal) through the projects CFTP-FCT Unit 777 (UIDB/00777/2020 and UIDP/00777/2020), PTDC/FIS-PAR/29436/2017, CERN/FIS-PAR/0008/2019 and CERN/FIS-PAR/0002/2021, which are partially funded through POCTI (FEDER), COMPETE, QREN and EU.

\providecommand{\noopsort}[1]{}\providecommand{\singleletter}[1]{#1}%
	
	\providecommand{\href}[2]{#2}\begingroup\raggedright\endgroup


\begin{thebibliography}{70}
%
%

\bibitem{Branco:1999fs}
G.~C.~Branco, L.~Lavoura and J.~P.~Silva,
\emph{{CP Violation,}}
Int. Ser. Monogr. Phys. \textbf{103} (1999), 1-536.

\bibitem{Workman:2022ynf} 
{\scshape Particle Data Group} collaboration, Workman, R. L. et~al.,
\emph{{Review of Particle Physics}},
{\emph{PTEP} {\bfseries 2022} (2022) 083C01} 

\bibitem{Zyla:2020zbs}
{\scshape Particle Data Group} collaboration, P.~Zyla et~al., \emph{{Review of Particle Physics}}, 
\href{https://doi.org/10.1093/ptep/ptaa104}{\emph{PTEP} {\bfseries 2020} (2020) 083C01}.


\bibitem{Grossman:2020qrp}
Y.~Grossman and J.~T.~Ruderman,
\emph{{CKM substructure}}, 
\href{https://doi.org/10.1007/JHEP01(2021)143}{\emph{JHEP} {\bfseries 01} (2021) 143}.
[\href{https://arxiv.org/abs/2007.12695}{{\ttfamily 2007.12695}}].
[arXiv:2007.12695 [hep-ph]].

\bibitem{Belfatto:2019swo}
B.~Belfatto, R.~Beradze and Z.~Berezhiani, \emph{{The CKM unitarity problem: A
trace of new physics at the TeV scale?}},
\href{https://doi.org/10.1140/epjc/s10052-020-7691-6}{\emph{Eur. Phys. J. C} {\bfseries 80} (2020) 149}
[\href{https://arxiv.org/abs/1906.02714}{{\ttfamily 1906.02714}}].

\bibitem{Seng:2018qru}
			C.~Y. Seng, M.~Gorchtein and M.~J. Ramsey-Musolf, \emph{{Dispersive evaluation
					of the inner radiative correction in neutron and nuclear $\beta$ decay}},
			\href{https://doi.org/10.1103/PhysRevD.100.013001}{\emph{Phys. Rev. D} {\bfseries 100} (2019) 013001}
			[\href{https://arxiv.org/abs/1812.03352}{{\ttfamily 1812.03352}}].
			
			\bibitem{Czarnecki:2019mwq}
			A.~Czarnecki, W.~J. Marciano and A.~Sirlin, \emph{{Radiative Corrections to
					Neutron and Nuclear Beta Decays Revisited}},
			\href{https://doi.org/10.1103/PhysRevD.100.073008}{\emph{Phys. Rev. D} {\bfseries 100} (2019) 073008}
			[\href{https://arxiv.org/abs/1907.06737}{{\ttfamily 1907.06737}}].
			
			
			\bibitem{Seng:2020wjq}
			C.-Y. Seng, X.~Feng, M.~Gorchtein and L.-C. Jin, \emph{{Joint lattice
					QCD\textendash{}dispersion theory analysis confirms the quark-mixing top-row
					unitarity deficit}},
			\href{https://doi.org/10.1103/PhysRevD.101.111301}{\emph{Phys. Rev. D} {\bfseries 101} (2020) 111301}
			[\href{https://arxiv.org/abs/2003.11264}{{\ttfamily 2003.11264}}].
			
			\bibitem{Hayen:2020cxh}
			L.~Hayen, \emph{{Standard Model $\mathcal{O}(\alpha)$ renormalization of $g_A$
					and its impact on new physics searches}},
			\href{https://doi.org/10.1103/PhysRevD.103.113001}{\emph{Phys. Rev. D} {\bfseries 103} (2021) 113001}
			\href{https://arxiv.org/abs/2010.07262}{{\ttfamily 2010.07262}}.
			
			
			\bibitem{Shiells:2020fqp}
			K.~Shiells, P.~G. Blunden and W.~Melnitchouk, \emph{{Electroweak axial
					structure functions and improved extraction of the $V_{ud}$ CKM matrix
					element}}, \href{https://doi.org/10.1103/PhysRevD.104.033003}{\emph{Phys. Rev. D} {\bfseries 104} (2021) 033003}
			\href{https://arxiv.org/abs/2012.01580}{{\ttfamily 2012.01580}}.
			
			\bibitem{Czarnecki:2004cw}
			A.~Czarnecki, W.~J. Marciano and A.~Sirlin, \emph{{Precision measurements and
					CKM unitarity}},
			\href{https://doi.org/10.1103/PhysRevD.70.093006}{\emph{Phys. Rev. D} {\bfseries 70} (2004) 093006}
			[\href{https://arxiv.org/abs/hep-ph/0406324}{{\ttfamily hep-ph/0406324}}].
			
			
			\bibitem{Coutinho:2019aiy}
			Antonio M. Coutinho, Andrea Crivellin, \emph{{Global Fit to Modified Neutrino Couplings}},
			\href{https://doi.org/10.1103/PhysRevLett.125.071802}{\emph{Phys. Rev. Lett.} {\bfseries 125} (2020) 071802}
			[\href{https://arxiv.org/abs/1912.08823}{{\ttfamily 1912.08823}}].
			
			\bibitem{Aoki:2021kgd}
			Y. Aoki {\it et al}, \emph{{FLAG Review 2021}},
			[\href{https://arxiv.org/abs/2111.09849}{{\ttfamily 2111.09849}}].
			
			\bibitem{Seng:2018yzq}
            C.-Y. Seng, M.~Gorchtein, H.~H. Patel and M.~J. Ramsey-Musolf, \emph{{Reduced
            Hadronic Uncertainty in the Determination of $V_{ud}$}},
            \href{https://doi.org/10.1103/PhysRevLett.121.241804}{\emph{Phys. Rev. Lett.} {\bfseries 121} (2018) 241804} 
            [\href{https://arxiv.org/abs/1807.10197}{{\ttfamily 1807.10197}}].



\bibitem{Buras_2010}
			A.~J.~Buras, B.~Duling, T.~Feldmann, T.~Heidsieck, C.~Promberger and S.~Recksiegel,
			\emph{{Patterns of Flavour Violation in the Presence of a Fourth Generation of Quarks and Leptons}},
			\href{https://doi.org/10.1007/JHEP09(2010)106}{JHEP \textbf{09} (2010), 106}
			[\href{https://arxiv.org/abs/1002.2126}{arXiv:1002.2126}].
			
\bibitem{Aguilar-Saavedra:2002phh}
J.~A.~Aguilar-Saavedra,
\emph{{Effects of mixing with quark singlets}},
\href{https://doi.org/10.1103/PhysRevD.69.099901}{Phys. Rev. D \textbf{67} (2003), 035003},
[erratum: Phys. Rev. D \textbf{69} (2004), 099901]
[\href{https://arxiv.org/abs/hep-ph/0210112}{hep-ph/0210112}].

\bibitem{Cacciapaglia_2012}
			G.~Cacciapaglia, A.~Deandrea, L.~Panizzi, N.~Gaur, D.~Harada and Y.~Okada,
			\emph{{Heavy Vector-like Top Partners at the LHC and flavour constraints}},
			\href{https://doi.org/10.1007/JHEP03(2012)070}{JHEP \textbf{03} (2012), 070}
			[\href{https://arxiv.org/abs/1108.6329}{arXiv:1108.6329}].

\bibitem{Botella_2017}
			Francisco J.Botella, G.~C.~Branco, Miguel Nebot, M.~N.~Rebelo, and J.~I.~Silva-Marcos.,
			\emph{{Vector-like quarks at the origin of light quark masses and mixing}},
			\href{https://doi.org/10.1140/epjc/s10052-017-4933-3}{Eur. Phys. J. C \textbf{77} (2017)  408}
			[\href{https://arxiv.org/abs/1610.03018}{arXiv:1610.03018}]
			
\bibitem{Botella_2017_a}
			Francisco J. Botella, Gustavo C. Branco and Miguel Nebot,
			\emph{{Singlet Heavy Fermions as the Origin of B Anomalies in Flavour Changing Neutral Currents}},
			[\href{https://arxiv.org/abs/1712.04470}{arXiv:1712.04470}].
			
\bibitem{Nardi_1996}
			Enrico Nardi, 
			\emph{{Top - charm flavor changing contributions to the effective bsZ vertex}},
			\href{https://doi.org/10.1016/0370-2693(95)01308-3}{Phys. Lett. B \textbf{365} (1996) 327}
			[\href{https://arxiv.org/abs/hep-ph/9509233}{hep-ph/9509233}]



\bibitem{Barenboim:1997pf}
			G.~Barenboim and F.~J. Botella, \emph{{Delta F=2 effective Lagrangian in
					theories with vector - like fermions}},
			\href{https://doi.org/10.1016/S0370-2693(98)00695-9}{\emph{Phys. Lett. B}{\bfseries 433} (1998) 385}
			[\href{https://arxiv.org/abs/hep-ph/9708209}{{\ttfamily hep-ph/9708209}}].
			
\bibitem{Balaji:2021lpr}
    S.~Balaji,
    \emph{{Asymmetry in flavour changing electromagnetic transitions of vector-like quarks,}}
	\href{https://doi.org/10.1007/JHEP05(2022)015}{JHEP \textbf{05} (2022), 015}
    [\href{https://arxiv.org/abs/2110.05473}{{\ttfamily hep-ph/2110.05473}}].
			
			
			\bibitem{Branco_1995}
G.~C.~Branco, P.~A.~Parada and M.~N.~Rebelo,
\emph{{D0 - anti-D0 mixing in the presence of isosinglet quarks}},
\href{https://doi.org/10.1103/PhysRevD.52.4217}{Phys. Rev. D \textbf{52} (1995), 4217-4222}
[\href{https://arxiv.org/abs/hep-ph/9501347}{hep-ph/9501347}].
			
			\bibitem{Botella:2021uxz}
F.~J.~Botella, G.~C.~Branco, M.~N.~Rebelo, J.~I.~Silva-Marcos and J.~Filipe Bastos,
\emph{{Decays of the heavy top and new insights on $\varepsilon _K$ in a one-VLQ minimal solution to the CKM unitarity problem,}}
\href{https://doi.org/10.1140/epjc/s10052-022-10299-9}{Eur. Phys. J. C \textbf{82} (2022) no.4, 360}
[\href{https://arXiv.org/abs/2111.15401}{arXiv:2111.15401}].

\bibitem{Branco:2021vhs}
G. C. Branco, J. T. Penedo, Pedro M. F. Pereira, M. N. Rebelo and J. I. Silva-Marcos, 
\href{https://doi.org/10.1007/JHEP07(2021)099}{\emph{{Addressing the CKM unitarity problem with a vector-like up quark}}, 
JHEP \textbf{07} (2021), 099}
[ \href{https://arxiv.org/abs/2103.13409}{{\ttfamily 2103.13409}}].

\bibitem{Belfatto:2021jhf}
			B.~Belfatto and Z.~Berezhiani,
			\href{https://doi.org/10.1007/JHEP10(2021)079}{\emph{{Are the CKM anomalies induced by vector-like quarks? Limits from flavor changing and Standard Model precision tests}}, 
				JHEP \textbf{10} (2021), 079}
			[\href{https://arXiv.org/abs/2103.05549}{{\ttfamily  2103.13409}}].

\bibitem{Amhis_2021}
Y.~S.~Amhis \textit{et al.} [HFLAV],
\emph{{Averages of b-hadron, c-hadron, and $\tau $-lepton properties as of 2018}},
\href{https://doi.org/10.1140/epjc/s10052-020-8156-7}{Eur. Phys. J. C \textbf{81} (2021) no.3, 226}
[\href{https://arxiv.org/abs/1909.12524}{arXiv:1909.12524}].

	\bibitem{Golowich_2009}
			E.~Golowich, J.~Hewett, S.~Pakvasa and A.~A.~Petrov,
			\emph{{Relating D0-anti-D0 Mixing and D0 ---\ensuremath{>} l+ l- with New Physics}},
			\href{https://doi.org/10.1103/PhysRevD.79.114030}{Phys. Rev. D \textbf{79} (2009), 114030}
			[\href{https://arxiv.org/abs/0903.2830}{arXiv:0903.2830}].
			
\bibitem{Sirunyan_2018}
A.~M.~Sirunyan \textit{et al.} [CMS],
\emph{{Search for vectorlike light-flavor quark partners in proton-proton collisions at $\sqrt s$ =8  TeV}},
\href{https://doi.org/10.1103/PhysRevD.97.072008}{Phys. Rev. D \textbf{97} (2018), 072008}
[\href{https://arxiv.org/abs/1708.02510}{arXiv:1708.02510} ].

\bibitem{Botella:1985gb}
			F.~Botella and L.-L. Chau, \emph{{Anticipating the Higher Generations of Quarks
					from Rephasing Invariance of the Mixing Matrix}},
			\href{https://doi.org/10.1016/0370-2693(86)91468-1}{\emph{Phys. Lett. B}
				{\bfseries 168} (1986) 97}.
\bibitem{Brod_2020}
J.~Brod, M.~Gorbahn and E.~Stamou,
\emph{{Standard-Model Prediction of $\varepsilon_K$ with Manifest Quark-Mixing Unitarity}},
\href{https://doi.org/10.1103/PhysRevLett.125.171803}{Phys. Rev. Lett. \textbf{125} (2020) no.17, 171803}
[\href{https://arxiv.org/abs/1911.06822}{arXiv:1911.06822}]
				
\bibitem{Inami:1980fz}
T.~Inami and C.~S.~Lim,
\emph{{Effects of Superheavy Quarks and Leptons in Low-Energy Weak Processes k(L) ---\ensuremath{>} mu anti-mu, K+ ---\ensuremath{>} pi+ Neutrino anti-neutrino and K0 \ensuremath{<}---\ensuremath{>} anti-K0}},
\href{https://doi.org/10.1143/PTP.65.297}{Prog. Theor. Phys. \textbf{65} (1981), 297}
[erratum: Prog. Theor. Phys. \textbf{65} (1981), 1772]
		
		
\end{thebibliography}
\end{document}